\documentstyle[11pt,moriond,epsfig]{article}

\def \Sleuth {{\sc sleuth}}
\def \met {{\,/\!\!\!\!E_{T}}}
\def \hse {{\small{hse}}}
\def \hser {{\small{hser}}}

\def \jj {\,2j}
\def \jjj {\,3j}
\def \jjjj {\,4j}
\def \jjjjj {\,5j}
\def \jjjjjj {\,6j}

\def \Nmin {{N_{\rm min}}} 
\def \scriptR {\mbox{${\cal R}$}}
\def \scriptP {\mbox{${\cal P}$}}
\def \gothicP {\tilde{\scriptP}}

\begin{document}
\vspace*{4cm}
\title{SLEUTH:  A QUASI-MODEL-INDEPENDENT SEARCH STRATEGY FOR NEW PHYSICS}

\author{ BRUCE KNUTESON }

\address{Enrico Fermi Institute, University of Chicago, 5640 S. Ellis Ave.,\\
Chicago, IL 60637\\
\vspace*{0.10truein}
\centerline{\footnotesize for the D\O\ collaboration}}

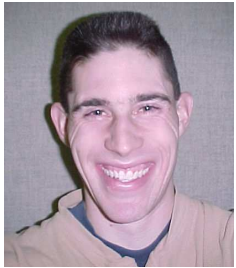
\maketitle\abstracts{
How can we search for new physics when we only vaguely know what it should look like?  How can we perform an unbiased yet data-driven search?  If we see apparently anomalous events in our data, how can we quantify their ``interestingness'' {\em a posteriori}?  We present an analysis strategy (\Sleuth) that simultaneously addresses each of these questions, and we demonstrate its application to over thirty exclusive final states in data collected by D\O\ in Run I of the Fermilab Tevatron.
}

\section{Motivation}

It is generally recognized that the standard model, an extremely successful description of the fundamental particles and their interactions, must be incomplete.  Although there is likely to be new physics beyond the current picture, the possibilities are sufficiently broad that the first hint could appear in any of many different guises.  This suggests the importance of performing searches that are as model-independent as possible.

Most recent searches for new physics have followed a well-defined set of steps:  first selecting a model to be tested against the standard model, then finding a measurable prediction of this model that differs as much as possible from the prediction of the standard model, and finally comparing the predictions to \mbox{data}.  This is clearly the procedure to follow for a small number of compelling candidate theories.  Unfortunately, the resources required to implement this procedure grow almost linearly with the number of theories.  Although broadly speaking there are currently only three models with internally consistent methods of electroweak symmetry breaking --- supersymmetry, strong dynamics, and theories incorporating large extra dimensions --- the number of specific models (and corresponding experimental signatures) is in the hundreds.  Of these many specific models, at most one is a correct description of nature.

In these proceedings we describe an explicit prescription 
for searching for the physics responsible for stabilizing electroweak symmetry breaking, in a manner that relies only upon what we are sure we know about electroweak symmetry breaking:  that its natural scale is on the order of the Higgs mass.  When we wish to emphasize the generality of the approach, we say that it is quasi-model-independent, where the ``quasi'' refers to the fact that the correct model of electroweak symmetry breaking should become manifest at the scale of several hundred GeV.

New sources of physics will in general lead to an excess over the expected background in some final state.  A general signature for new physics is therefore a region of variable space in which the probability for the background to fluctuate up to or above the number of observed events is small.  Because the mass scale of electroweak symmetry breaking is larger than the mass scale of most standard model backgrounds, we expect this excess to populate regions of high transverse momentum ($p_T$).  The method we will describe involves a systematic search for such excesses.  Although motivated by the problem of electroweak symmetry breaking, this method is generally sensitive to any new high $p_T$ physics.

\section{\Sleuth}

\Sleuth,\cite{SherlockPRD1,SherlockPRD2,SherlockPRL,KnutesonThesis} a quasi-model-independent prescription for searching for high $p_T$ physics beyond the standard model, has three components:  
the definitions of physical objects and exclusive final states;
the choice of variables relevant for each final state; and
an algorithm that systematically hunts for an excess in the space of those variables, and quantifies the likelihood of any excess found.
We consider each in turn.

\subsection{Final states}  
The data are partitioned into exclusive final states using standard criteria that identify isolated and energetic electrons ($e$), muons ($\mu$), and photons ($\gamma$), as well as jets ($j$), missing transverse energy ($\met$), and the presence of $W$ and $Z$ bosons.  We expect the first sign of new physics to appear in one of these final states, but which final state it will be is anyone's guess.  We analyze each of these final states independently.

\subsection{Variables}
For each exclusive final state, we consider a small set of variables summarized in Table~\ref{tbl:VariableRules}.\cite{SherlockPRD1,SherlockPRL}

\begin{table}[h]
\caption{A quasi-model-independently motivated list of interesting variables for any final state.  The set of variables to consider for any exclusive channel is the union of the variables in the second column for each row that pertains to that final state.}
\label{tbl:VariableRules}
\vspace{0.4cm}
\begin{center}
\begin{tabular}{|c|c|}
\hline
If the final state includes & then consider the variable \\ \hline
$\met$ & $\met$ \\ 
one or more charged leptons & $\sum{p_T^\ell}$ \\ 
one or more electroweak bosons & $\sum{p_T^{\gamma/W/Z}}$ \\ 
one or more jets & $\sum{p_T^j}$ \\ 
\hline
\end{tabular}
\end{center}
\end{table}

\subsection{Algorithm}
The \Sleuth\ algorithm requires as input a data sample, a set of events modeling each background process $i$, and the number of background events $\hat{b}_i \pm \delta \hat{b}_i$ from each background process expected in the data sample.  From these we determine the region $\scriptR$ of greatest excess and quantify the degree $\scriptP$ to which that excess is interesting.  The algorithm itself, applied to each individual final state, consists of seven steps:  

\begin{enumerate}

\item We construct a mapping from the $d$-dimensional variable space defined by Table~\ref{tbl:VariableRules} into the $d$-dimensional unit box (i.e., $[0,1]^d$) that flattens the total background distribution.  We use this to map the data into the unit box.  

\item We define a ``region'' $R$ about a set of $N$ data points to be the volume within the unit box closer to one of the data points in the set than to any of the other data points in the sample.  The arrangement of data points themselves thus determines the regions.  A region containing $N$ data points is called an $N$-region.

\item Each region contains an expected number of background events $\hat{b}_R$, numerically equal to the volume of the region $\times$ the total number of background events expected, and an associated systematic error $\delta\hat{b}_R$, which varies within the unit box according to the systematic errors assigned to each contribution to the background estimate.  We can therefore compute the probability $p_N^R$ that the background in the region fluctuates up to or beyond the observed number of events.  This probability is the first measure of the degree of interest of a particular region.

\item The rigorous definition of regions reduces the number of candidate regions from infinity to $\approx 2^{N_{\rm data}}$.  Imposing explicit criteria on the regions that the algorithm is allowed to consider further reduces the number of candidate regions.  We apply geometric criteria that favor high values in at least one dimension of the unit box, and we limit the number of events in a region to fifty.  The number of remaining candidate regions is still sufficiently large that an exhaustive search is impractical, and a heuristic is employed to search for regions of excess.  In the course of this search, the $N$-region $\scriptR_N$ for which $p_N^R$ is minimum is determined for each $N$, and $p_N = \min_R{(p_N^R)}$ is noted.

\item In any reasonably-sized data set, there will always be regions in which the probability for $b_R$ to fluctuate up to or above the observed number of events is small.  We determine the fraction $P_N$ of {\em hypothetical similar experiments} (\hse's) in which $p_N$ found for the \hse\ is smaller than $p_N$ observed in the data by generating random events drawn from the background distribution and computing $p_N$ by following steps 1--4. 

\item We define $P$ and $\Nmin$ by $P=P_\Nmin=\min_N{(P_N)}$, and identify $\scriptR = \scriptR_\Nmin$ as the most interesting region in this final state.  

\item We use a second ensemble of \hse's to determine the fraction $\scriptP$ of \hse's in which $P$ found in the \hse\ is smaller than $P$ observed in the data.  The most important output of the algorithm is this single number $\scriptP$, which may loosely be said to be the ``fraction of hypothetical similar experiments in which you would see an excess as interesting as what you actually saw in the data.''  $\scriptP$ takes on values between zero and unity, with values close to zero indicating a possible hint of new physics.  The computation of $\scriptP$ rigorously takes into account the many regions that have been considered within this final state.

\end{enumerate}

The smallest $\scriptP$ found in the many different final states considered ($\scriptP_{\rm min}$) determines $\gothicP$, the ``fraction of {\em hypothetical similar experimental runs} (\hser's) that would have produced an excess as interesting as actually observed in the data,'' where an \hser\ consists of one \hse\ for each final state considered.  $\gothicP$ is calculated by simulating an ensemble of hypothetical similar experimental runs, and noting the fraction of these \hser's in which the smallest $\scriptP$ found is smaller than the smallest $\scriptP$ observed in the data.  Because $\gothicP$ depends only on the single final state that defines $\scriptP_{\rm min}$, correlations among final states may be neglected in this calculation.  Like $\scriptP$, $\gothicP$ takes on values between zero and unity, and the potential presence of new high $p_T$ physics would be indicated by finding $\gothicP$ to be small.  The difference between $\gothicP$ and $\scriptP$ is that in computing $\gothicP$ we account for the many final states that have been considered.

\section{Results}

\Sleuth's performance on representative signatures has been studied.\cite{SherlockPRD1,SherlockPRD2,SherlockPRL,KnutesonThesis}  When ignorance of $t\bar{t}$ is feigned in the $e\mu X$ final states, we find $\scriptP_{e\mu\met \jj} = 1.9\sigma$ in D\O\ data, correctly suggesting the presence of $t\bar{t}$.  Feigning ignorance of $t\bar{t}$ in the $W$+jets-like final states, we find $\scriptP_{\rm min}>3\sigma$ in 30\% of an ensemble of mock experimental runs on the final states $W\jjj$, $W\jjjj$, $W\jjjjj$, and $W\jjjjjj$.  Dedicated searches for the top quark in these channels\cite{topCrossSection} yield an excess of $2.75\sigma$ in $e\mu\met\jj$, $2.6\sigma$ in $W\jjjj(nj)$ with no $b$-tag, and $3.6\sigma$ in $W\jjj(nj)$ with a $b$-tag.  We see that \Sleuth\ performs surprisingly well despite being denied $b$-tagging information and any information of the properties of the signal it is supposed to be looking for.  These and similar studies performed in other final states suggest that \Sleuth\ is sensitive to a variety of new physics signatures.

We have applied \Sleuth\ to over thirty exclusive final states at D\O, determining $\scriptP$ for each.  Upon taking into account the many final states (both populated and unpopulated) that are considered, we find $\gothicP$=0.89, implying that 89\% of an ensemble of hypothetical similar experimental runs would have produced a final state with a candidate signal more interesting than the most interesting observed in these data.  

\section{Conclusions}

We have applied \Sleuth\ to search for new high $p_T$ physics in data spanning over thirty exclusive final states collected by the D\O\ experiment during Run I of the Fermilab Tevatron.  A quasi-model-independent, systematic search of these data has produced no evidence of physics beyond the standard model.  The strategy proposed here may prove useful in future searches for new phenomena.

\section*{Acknowledgments}
The author's attendance at Moriond was supported in part by Boaz Klima and funds from the National Science Foundation.

\end{document}